\documentclass[reprint,aps,pra,floatfix,twocolumn]{revtex4-1}
\usepackage{graphicx}
\usepackage{amssymb}
\usepackage{dsfont}
\usepackage{bm}
\usepackage{epstopdf}

\begin{document}

\title{Continuous deformations of the Grover walk preserving localization}

\author{M. \v Stefa\v n\'ak\email[correspondence to:]{martin.stefanak@fjfi.cvut.cz}}
\affiliation{Department of Physics, Faculty of Nuclear Sciences and Physical Engineering, Czech Technical University in Prague, B\v
rehov\'a 7, 115 19 Praha 1 - Star\'e M\v{e}sto, Czech Republic}

\author{I. Bezd\v ekov\'a}
\affiliation{Department of Physics, Faculty of Nuclear Sciences and Physical Engineering, Czech Technical University in Prague, B\v
rehov\'a 7, 115 19 Praha 1 - Star\'e M\v{e}sto, Czech Republic}

\author{I. Jex}
\affiliation{Department of Physics, Faculty of Nuclear Sciences and Physical Engineering, Czech Technical University in Prague, B\v
rehov\'a 7, 115 19 Praha 1 - Star\'e M\v{e}sto, Czech Republic}

\date{\today}

\begin{abstract}

The three-state Grover walk on a line exhibits the localization effect characterized by a non-vanishing probability of the particle to stay at the origin. We present two continuous deformations of the Grover walk which preserve its localization nature. The resulting quantum walks differ in the rate at which they spread through the lattice. The velocities of the left and right-traveling probability peaks are given by the maximum of the group velocity. We find the explicit form of peak velocities in dependence on the coin parameter. Our results show that localization of the quantum walk is not a singular property of an isolated coin operator but can be found for entire families of coins.

\end{abstract}

\maketitle


\section{Introduction}

Quantum walks have been introduced by Aharonov et al. \cite{aharonov} as a generalization of a classical random walk \cite{hughes} to a unitary discrete-time evolution of a quantum particle. The particle moves on a graph or a lattice in discrete time-steps according to its internal degree of freedom, which is usually refer to as coin. In analogy with the coin tossing which tells the classical particle where to go the state of the coin is altered by the coin operator before the displacement itself. However, in a quantum walk each trajectory is assigned a certain probability amplitude and not a probability. Different trajectories interfere which leads to a ballistic spreading of a probability density of a quantum particle. Indeed, quantum walk can be considered as a wave phenomena \cite{knight}. This analogy allowed to adopt and develop a number of concepts used in wave propagation in material media for quantum walks. For instance, Kempf and Portugal defined the hitting time based on the concept of group velocity \cite{kempf:group:vel}.

The application of quantum walks for quantum information processing have been proposed \cite{kempe}. In particular, one can use the quantum walk to implement the quantum search algorithm \cite{shenvi:2003}. The performance of the search algorithm crucially depends on the choice of the coin operator \cite{ambainis:search}. A review of quantum walk based algorithms can be found in \cite{santha:2008}.

For a two-state walk on a line the coin operator is given by a $U(2)$ matrix. Nevertheless, it has been shown \cite{tregenna} that it is sufficient to consider the one-parameter family of coins
$$
C(\rho) = \left(
            \begin{array}{cc}
              \rho & \sqrt{1-\rho^2} \\
               \sqrt{1-\rho^2} & -\rho \\
            \end{array}
          \right),
$$
with $0\leq\rho\leq 1$. The phases in a general $U(2)$ matrix turn out to be either irrelevant for the quantum walk evolution or can be compensated by the choice of the initial state. The choice of $\rho=1/\sqrt{2}$ corresponds to the most studied case of the Hadamard walk \cite{ambainis:line}. The coin parameter determines the rate at which the walk spreads through the lattice. For unbiased walks the peaks of the probability density propagates with constant velocity $\pm\rho$. This also illustrates the ballistic nature of a quantum walk. Biasing the walk by allowing the particle to make longer jumps in one direction speeds up one of the peaks and slows down the other \cite{stef:njp}. This has a crucial impact on the recurrence properties of the quantum walk \cite{stef:prl}. The understanding of recurrence requires the knowledge of the asymptotic properties of a quantum walk. For two-state quantum walks these characteristics can be obtained from the limit theorems derived by Konno \cite{konno:limit:2002,konno:2005b}. Grimmett et al. \cite{Grimmett} have extended the weak limit theorems to higher-dimensional quantum walks. For a review of asymptotic methods in quantum walks see \cite{konno:book}.

Allowing the particle to stay at its actual position we have to extend the coin to a $U(3)$ matrix. The resulting three-state quantum walks lead to dynamics which cannot occur in the two-state walk. As an example, the intriguing effect of localization has been found in the three-state Grover walk on a line \cite{konno:loc:2005,konno:loc:2005b}. Here the particle has a non-vanishing probability to stay at the origin. However, the localization effect is sensitive to the dimensionality of the lattice. There is no localization in the three-state Grover walk on a triangular lattice \cite{kollar}. Nevertheless, localization is not limited to quantum walks which allow the particle to stand still. The Grover walk on a 2D square lattice represents such an example \cite{konno:loc:2004}.

We note that there are two types of localization in the context of quantum walks. The one we have just discussed and which we will focus on in the present paper is inherent to certain quantum walks without any perturbations. It stems from the fact that the unitary propagator of the walk has a non-empty point spectrum. As the wave packet spreads it overlaps with the corresponding bound states which results in partial trapping of the particle in the vicinity of the origin. The second kind is the Anderson localization which arises e.g. from static phase disorder \cite{yin:qw:loc} or spatial coin inhomogeneity \cite{konno:loc:2010}. This dynamical localization was experimentally observed in the photonic implementation of quantum walk on a line \cite{and:qw:loc}. For a comprehensive mathematical description of this effect we refer to the literature \cite{joye:qw:loc,ahlbrecht:qw:loc}.

In contrast to the two-state walk the properties of a three-state walk with a general $U(3)$ coin operator are not fully understood. The present paper is a step in classification of the three-state quantum walks. By deforming the Grover matrix we find two families of coins which lead to a localizing quantum walk. The first one-parameter family of coins is based on the variation of the spectrum of the Grover matrix. Another one-parameter family of coin operators is obtained by modifying the eigenvectors of the Grover matrix. In both cases we show how does the coin parameter determine the rate of spreading of the corresponding quantum walk by calculating the explicit form of the peak velocities. While for the first family of walks the peak velocities can only decrease when compared to the Grover walk, the second deformation allows to increase them to the maximum possible value.

Our manuscript is organized as follows: In Section~\ref{sec2} we briefly review the properties of the three-state Grover walk on a line following the Fourier analysis. We determine the peak velocities of the Grover walk by applying the stationary phase approximation in Section~\ref{sec3}. In Section~\ref{sec4} we introduce two deformations of the Grover walk which preserves its localization nature and analyze their peak velocities. We conclude and present an outlook in Section~\ref{sec5}.


\section{Three-state Grover walk on a line}
\label{sec2}

Let us first review the three-state Grover walk on a line \cite{konno:loc:2005,konno:loc:2005b}. The Hilbert space of the particle is given by the tensor product
$$
\mathcal{H} = \mathcal{H}_P \otimes \mathcal{H}_C
$$
of the position space
$$
\mathcal{H}_P = \rm{Span}\left\{|m\rangle,m\in\mathds{Z}\right\}
$$
and the coin space $\mathcal{H}_C$. In each step the particle has three possibilities - it can move to the left or right or stay at its current location. To each of these options we assign a vector of the standard basis of the coin space $\mathcal{H}_C$, i.e. the coin space is three-dimensional
$$
\mathcal{H}_C = \mathds{C}^3 = \rm{Span}\left\{|L\rangle,|S\rangle,|R\rangle\right\}.
$$
A single step of the quantum walk is realized by the propagator $U$ given by
$$
U = S\cdot (I_P\otimes C),
$$
where $S$ is the conditional step operator, $I_P$ denotes the identity on the position space and $C$ is the coin operator. For our three-state walk the conditional step operator $S$ has the following form
\begin{eqnarray}
\nonumber S = \sum_{m=-\infty}^{+\infty} & & \left(|m-1\rangle\langle m|\otimes |L\rangle\langle L| + |m\rangle\langle m|\otimes |S\rangle\langle S|\frac{}{}\right.\\
\nonumber & &\left. + |m+1\rangle\langle m|\otimes |R\rangle\langle R|\frac{}{}\right).
\end{eqnarray}
As the coin operator we choose the $3\times 3$ Grover matrix
$$
C = C_G = \frac{1}{3}\left(
                       \begin{array}{ccc}
                         -1 & 2 & 2 \\
                         2 & -1 & 2 \\
                         2 & 2 & -1 \\
                       \end{array}
                     \right).
$$

The state of the particle after $t$ steps is given by the successive application of the unitary propagator on the initial state
\begin{eqnarray}
\label{time:evol:x}
\nonumber |\psi(t)\rangle = \sum_{m} & & |m\rangle\left(\psi_L(m,t)|L\rangle + \psi_S(m,t)|S\rangle \frac{}{}\right. \\
& & \left. + \psi_R(m,t)|R\rangle\frac{}{}\right) = U^t |\psi(0)\rangle.
\end{eqnarray}
The probability distribution of the particle's position after $t$ steps of quantum walk is obtained by tracing out the coin degree of freedom
\begin{eqnarray}
\nonumber p(m,t)& = & |\psi_L(m,t)|^2 + |\psi_S(m,t)|^2 + |\psi_R(m,t)|^2 \\
\nonumber & = & ||\psi(m,t)||^2.
\end{eqnarray}
Here we have introduced the vector of probability amplitudes
$$
\psi(m,t) = \left(\psi_L(m,t),\psi_S(m,t),\psi_R(m,t)\right)^T.
$$
Since the walk we consider is translationally invariant the time evolution equation (\ref{time:evol:x}) greatly simplifies using the Fourier transformation
\begin{equation}
\label{FT}
\tilde{\psi}(k,t) = \sum_{m=-\infty}^{+\infty} e^{i m k} \psi(m,t),
\end{equation}
where the momentum $k$ ranges from $0$ to $2\pi$. Indeed, applying the Fourier transformation (\ref{FT}) to the time evolution equation (\ref{time:evol:x}), we find
\begin{equation}
\label{time:evol:p}
\tilde{\psi}(k,t) = \tilde{U}(k) \tilde{\psi}(k,t-1) = \tilde{U}^t(k) \tilde{\psi}(k,0).
\end{equation}
Here $\tilde{\psi}(k,0)$ denotes the Fourier transformation of the initial state of the particle. It equals the initial coin state $\psi_C$ of the particle provided that it starts the walk from the origin. The momentum representation of the time evolution operator $\tilde{U}(k)$ is given by
\begin{equation}
\label{U:k}
\tilde{U}(k) = \rm{Diag}\left(e^{-i k},1,e^{i k}\right)\cdot C_G.
\end{equation}
The time evolution equation in the momentum representation (\ref{time:evol:p}) is readily solved by diagonalizing the propagator (\ref{U:k}). We express the eigenvalues of $\tilde{U}(k)$ in the form $\lambda_j = \exp(i\omega_j(k))$ and denote the corresponding eigenvectors by $v_j(k)$. For the three-state Grover walk the phases read
\begin{eqnarray}
\label{phase:grover}
\nonumber \omega_{1,2}(k) & = & \pm \arccos\left(-\frac{1}{3}(2+\cos{k})\right),\\
\omega_3(k) & = & 0.
\end{eqnarray}
Since the phase $\omega_3$ vanishes we find that the corresponding eigenvalue $\lambda_3$ equals 1 independent of $k$. In other words, the propagator of the Grover walk has a non-empty point spectrum. This leads to the localization effect \cite{konno:loc:2005,konno:loc:2005b}. Finally, the solution of the time evolution equation in the momentum representation (\ref{time:evol:p}) has the form
$$
\tilde{\psi}(k,t) = \sum_{j=1}^3 e^{i\omega_j(k) t} \left(v_j(k),\psi_C\right)v_j(k).
$$
After the inverse Fourier transformation we obtain the solution in the position representation
\begin{equation}
\label{sol:x}
\psi(m,t) = \sum_{j=1}^3 \int\limits_{0}^{2\pi} \frac{dk}{2\pi}e^{i\left(\omega_j(k) - \frac{m}{t} k\right)t} \left(v_j(k),\psi_C\right)v_j(k).
\end{equation}


\section{Peak velocity of the Grover walk}
\label{sec3}

Let us now determine the rate at which the three-state Grover walk spreads through the lattice. We employ the stationary phase approximation \cite{statphase} which determines the behavior of the amplitude (\ref{sol:x}) for $t\rightarrow +\infty$. Accordingly, the rate of the decay is given by the order of the stationary points of the phase
\begin{equation}
\label{phase:mod}
\tilde{\omega}_j(k) \equiv \omega_j(k) - \frac{m}{t} k.
\end{equation}
The peak corresponds to the stationary point of the second order - both the first and the second derivatives of the phase (\ref{phase:mod}) with respect to $k$ vanish. Thus we have to solve a set of equations
\begin{eqnarray}
\label{phase:der}
\nonumber \frac{d\tilde{\omega}_j}{dk} = \frac{d\omega_j}{dk} - \frac{m}{t} = 0,\\
\frac{d^2\tilde{\omega}_j}{dk^2} = \frac{d^2\omega_j}{dk^2} = 0,
\end{eqnarray}
for $k$ and $m$. Assume that $k_0$ satisfies the second equation in (\ref{phase:der}). From the first equation in (\ref{phase:der}) we find that the position of the peak after $t$ steps is
$$
m = \left.\frac{d\omega_j}{dk}\right|_{k_0} t.
$$
The peak thus propagates with constant velocity which is given by $\left.\frac{d\omega_j}{dk}\right|_{k_0}$.

We find that there is a simple analogy with wave theory. Indeed, consider $k$ as wavenumber and $\omega_j(k)$ as frequency. Equations (\ref{phase:grover}) represent the dispersion relations. Taking the derivative with respect to $k$ we obtain the group velocity \cite{kempf:group:vel}. The wavefront, i.e. the peak in the probability distribution, propagates with the maximal group velocity.

Let us specify the results for the three-state Grover walk. From the explicit form of the dispersion relations (\ref{phase:grover}) we find that the second equation in (\ref{phase:der}) reads
$$
\frac{d^2\omega_{1,2}}{dk^2} = \pm 2 \sqrt{\frac{1-\cos{k}}{(5+\cos{k})^3}} = 0.
$$
This relation is satisfied for $k_0 = 0$. Evaluating the first derivative of $\omega_j(k)$ at this point we obtain the velocities of the left and right-traveling peaks
\begin{eqnarray}
\nonumber v_R & = & \lim\limits_{k\rightarrow 0^+}\frac{d\omega_2}{dk} = \frac{1}{\sqrt{3}},\\
v_L & = & \lim\limits_{k\rightarrow 0^+}\frac{d\omega_1}{dk} = -\frac{1}{\sqrt{3}}.
\end{eqnarray}
Note that from the constant phase $\omega_3 \equiv 0$ one immediately obtains $v_S = 0$. Indeed, the constant eigenvalue results in the central peak of the probability distribution which does not propagate.

\begin{figure}[h]
\begin{center}
\includegraphics[width=0.45\textwidth]{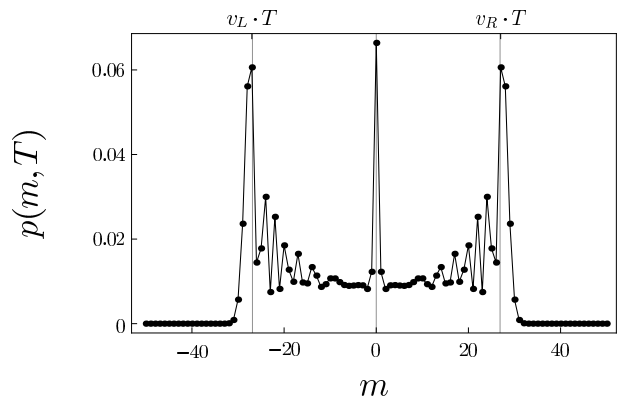}
\caption{The probability distribution of the three-state Grover walk after $T = 50$ steps. The initial coin state of the walk was $\psi_C = \frac{1}{\sqrt{3}}\left(1,-1,1\right)$. One can clearly identify three dominant peaks in the probability distribution. The peak at the origin corresponds to the localization nature of the Grover walk and does not propagate. The peaks on the sides travel with constant velocities $v_{L,R} = \pm\frac{1}{\sqrt{3}}$. The grid-lines corresponding to $T v_{L,R}\approx \pm 29$ coincides with the positions of the peaks obtained from the numerical simulation.}
\label{fig1}
\end{center}
\end{figure}

To illustrate our results we plot in Figure~\ref{fig1} the probability distribution of the three-state Grover walk after $T=50$ steps. The probability distribution contains three dominant peaks. Their positions are determined by the velocities $v_{L,R}$ and $v_S$.

\section{Deformations of the Grover walk}
\label{sec4}

We begin with the spectral decomposition of the Grover coin. Consider the following orthonormal basis
\begin{eqnarray}
\label{eigenvec}
\nonumber v_1 & = & \frac{1}{\sqrt{6}}\left(1,-2,1\right)^T,\\
\nonumber v_2 & = & \frac{1}{\sqrt{2}}\left(1,0,-1\right)^T,\\
 v_3 & = & \frac{1}{\sqrt{3}}\left(1,1,1\right)^T,
\end{eqnarray}
formed by the eigenvectors of the Grover coin with eigenvalues $\lambda_1 = \lambda_2 = -1$ and $\lambda_3 = 1$. The Grover coin can be thus decomposed in the form
\begin{equation}
\label{decomp:grover}
C_G = \sum_{j=1}^3 \lambda_j P_j = -P_1 - P_2 + P_3,
\end{equation}
where $P_j$ is the projection on the subspace spanned by the corresponding eigenvector.

In the following we introduce two deformations of the three-state Grover walk which preserves its localization nature. First, we modify the eigenvalues of the coin and keep the eigenvectors constant. Second, we leave the spectrum unchanged and deform the eigenvectors.

\subsection{Deforming the eigenvalues}
\label{sec4:1}

The Grover matrix is very symmetric - it is invariant under all permutations of the basis states. Hence, we can diagonalize it together with any permutation matrix. The eigenvectors of the Grover matrix presented in Eq. (\ref{eigenvec}) are chosen in such a way that they are also eigenvectors of the permutation matrix
$$
\Pi = \left(
               \begin{array}{ccc}
                 0 & 0 & 1 \\
                 0 & 1 & 0 \\
                 1 & 0 & 0 \\
               \end{array}
             \right).
$$
The corresponding eigenvalues are $\mu_1=\mu_3=1$ and $\mu_2= -1$. The permutation $\Pi$ interchanges the $|L\rangle$ and $|R\rangle$ coin states and preserves the $|S\rangle$ state. Using this matrix as a coin for a three-state quantum walk results in a trivial evolution - the particle either stays at the origin, or jumps to the left or right but immediately returns back in the next step. Such a walk does not spread through the lattice and the velocities vanish.

Notice that the eigenvector $v_2$ corresponds to the same eigenvalue $\lambda_2=\mu_2=-1$ for both the Grover and the permutation matrix $\Pi$. The same applies to $v_3$ since $\lambda_3=\mu_3 = 1$. However, for $v_1$ we have $\lambda_1= -1$ and $\mu_1 = 1$. This suggest to introduce a phase factor in front of the projector $P_1$ in the spectral decomposition of the Grover coin (\ref{decomp:grover}). In this way we can continuously change from the Grover coin to the permutation matrix $\Pi$. We thus arrive at the following one-parameter family of coin operators
\begin{eqnarray}
\label{C:phi}
\nonumber C_1(\varphi) & = & -e^{2i\varphi} P_1 - P_2 + P_3 \\
\nonumber & = & \frac{1}{6}\left(
                                                 \begin{array}{ccc}
                                                   -1-e^{2i\varphi} & 2(1+e^{2i\varphi}) & 5-e^{2i\varphi} \\
                                                   2(1+e^{2i\varphi}) & 2(1-2e^{2i\varphi}) & 2(1+e^{2i\varphi}) \\
                                                   5-e^{2i\varphi} & 2(1+e^{2i\varphi}) & -1-e^{2i\varphi} \\
                                                 \end{array}
                                               \right).\\
\end{eqnarray}
The factor of 2 in the exponent was included for convenience. We show that the family of three-state quantum walks with the coin operator (\ref{C:phi}) posses the localization property of the Grover walk. The phase parameter $\varphi$ determines the rate of spreading of the probability distribution through the lattice.

The dispersion relations for the one-parameter family of quantum walks with the coin operator (\ref{C:phi}) are
\begin{eqnarray}
\nonumber \omega_{1,2}(k,\varphi) & = & \varphi \pm \arccos\left(-\frac{1}{3} (2+\cos{k}) \cos{\varphi}\right),\\
\nonumber \omega_3(k,\varphi)& = & 0.
\end{eqnarray}
As for the original Grover walk we find that one frequency is independent of $k$. This ensures that the localization effect is preserved.

Let us now determine the velocities of the peaks, i.e. the maximal group velocity. The second derivatives of the frequencies $\omega_{1,2}(k,\varphi)$
$$
\frac{\partial^2\omega_{1,2}}{\partial k^2} = \mp\frac{9 \cos{k} -
 \cos^2{\varphi} (2 + 5 \cos{k} + 2\cos^2{k})}{(9 -
  \cos^2{\varphi}(2 + \cos{k})^2)^\frac{3}{2}}\cos{\varphi}
$$
vanish for
\begin{eqnarray}
\nonumber k_0 = \arccos & & \left(\frac{1}{4\cos^2{\varphi}} \left(9-5 \cos^2{\varphi}- \right.\right. \\
& &
\nonumber \left.\left.3 \sqrt{9-10\cos^2{\varphi}+\cos^4{\varphi}}\right)\right).
\end{eqnarray}
Evaluating the group velocities $\frac{\partial\omega_{1,2}}{\partial k}$ at the stationary point $k_0$ we obtain the velocities of the left and right-going peaks
\begin{eqnarray}
\label{vel:C:phi}
\nonumber v_{R}(\varphi) & = & \left.\frac{\partial \omega_2}{\partial k}\right|_{k_0} = \frac{1}{\sqrt{6}} \sqrt{3-\cos^2{\varphi}- \sin{\varphi}\sqrt{9-\cos^2{\varphi}}},\\
v_L(\varphi) & = & \left.\frac{\partial \omega_1}{\partial k}\right|_{k_0} = -v_R(\varphi).
\end{eqnarray}

\begin{figure}[h]
\begin{center}
\includegraphics[width=0.45\textwidth]{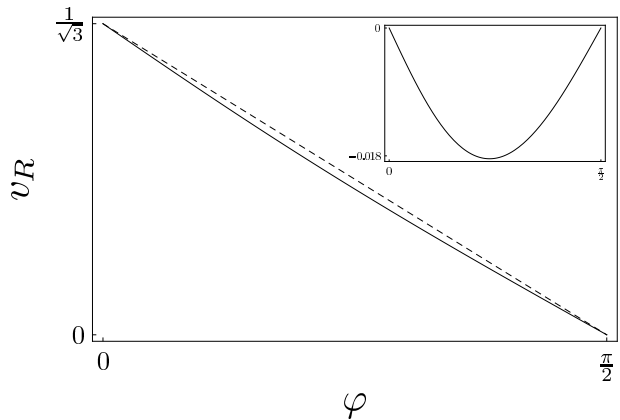}
\caption{The velocity of the $v_R(\varphi)$ for the one-parameter family of quantum walks defined by the coin operator $C_G(\varphi)$ (\ref{C:phi}). Despite the rather complicated formula (\ref{vel:C:phi}) for $v_R(\varphi)$ we see that it decreases almost linearly with $\varphi$. The dashed curve corresponds to the straight line $\frac{1}{\sqrt{3}}(1 - 2\varphi/\pi)$. The inset shows the difference of the two curves.}
\label{fig2}
\end{center}
\end{figure}

\begin{figure}[h]
\begin{center}
\includegraphics[width=0.45\textwidth]{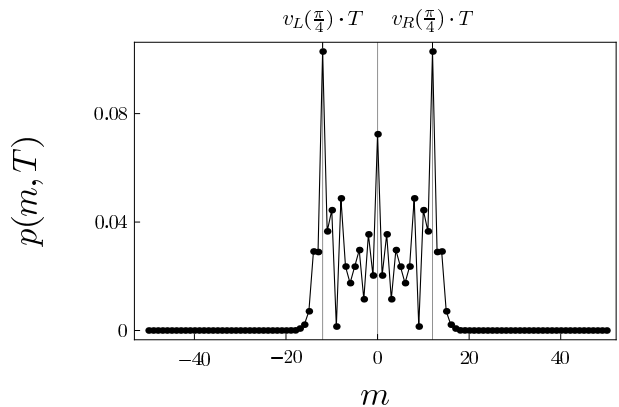}
\caption{The probability distribution of the three-state walk with the coin operator $C_1(\pi/4)$ after $T = 50$ steps. As for the Grover walk we have chosen the initial coin state according to $\psi_C = \frac{1}{\sqrt{3}}\left(1,-1,1\right)$. When compared with Figure~\ref{fig1} we find that the spreading of the probability distribution is much slower than for the Grover walk. Indeed, for $\varphi=\pi/4$ the peak velocities drops down to $v_{L,R}(\pi/4) \approx \pm 0.27$, which is less than a half of the velocities for the Grover walk.}
\label{fig3}
\end{center}
\end{figure}

We illustrate our results in Figures~\ref{fig2} and \ref{fig3}. In Figure~\ref{fig2} we display the velocity $v_R(\varphi)$ as a function of the coin parameter $\varphi$. It turns out that the dependence is almost linear
$$
v_R(\varphi) \approx \frac{1}{\sqrt{3}}\left(1-\frac{2\varphi}{\pi}\right).
$$
The inset shows the variation of $v_R(\varphi)$ from this linear approximation. We find that from the one-parameter family of quantum walks with the coin operator $C_G(\varphi)$ the Grover walk is the fastest one, as $v_R(\varphi)$ attains the maximal value $\frac{1}{\sqrt{3}}$ for $\varphi=0$. With increasing $\varphi$ the velocity of the right peak drops down and it becomes zero for $\varphi=\pi/2$.

In Figure~\ref{fig3} we show the probability distribution of the generalized three-state localizing walk with the parameter $\varphi=\pi/4$ after $T=50$ steps. In comparison with the original Grover walk displayed in Figure~\ref{fig1} we find that the distribution spreads much slower.

\subsection{Deforming the eigenvectors}
\label{sec4:2}

Our second approach to the deformation of the Grover walk is inspired by the work of Watabe et al. \cite{konno:loc:2008}. The authors have studied a one-parameter family of 2D four-state quantum walks which contained the Grover walk. This set of quantum walks also preserves the localization effect. The particular property of the corresponding one-parameter set of $4\times 4$ coin operators is that they have the same spectrum as the Grover matrix. We show that this feature can be employed to construct a similar set of $3\times 3$ coins.

Let us first consider two rather trivial coin operators which have the same spectrum as the Grover matrix and which also preserve localization of the corresponding quantum walk. One of such matrices is similar to the permutation matrix $\Pi$ introduced in the previous section, namely
$$
C = \left(
\begin{array}{ccc}
 0 & 0 & 1 \\
 0 & -1 & 0 \\
 1 & 0 & 0
\end{array}
\right).
$$
We have only changed the sign of the diagonal element which ensures that $C$ has the same spectrum as the Grover matrix. Nevertheless, the corresponding quantum walk is the same as the walk with the permutation coin $\Pi$. The walk is trivially localizing and the peak velocities equal zero. The second coin operator we consider is given by
$$
C' = \left(
\begin{array}{ccc}
 -1 & 0 & 0 \\
 0 & 1 & 0 \\
 0 & 0 & -1
\end{array}
\right).
$$
The dynamics of the resulting quantum walk is simple. The $|S\rangle$ component of the initial state remains at the origin which corresponds to localization. The $|L\rangle$ ($|R\rangle$) component moves in every step to the left (right). After $t$ steps the particle can be found only on three lattice points - either $m=0$ or $m=\pm t$. In contrast to the walk driven by the coin $C$ the walk with coin $C'$ spreads through the lattice with the maximal possible peak velocities $v_{L,R}=\pm 1$.

In order to connect the Grover matrix and the matrices $C$ and $C'$ we examine their eigenvectors. The eigenvectors of the Grover matrix were given in (\ref{eigenvec}). The eigenvectors of $C$ are
\begin{eqnarray}
\nonumber u_1 & = & \left(0,-1,0\right)^T,\\
\nonumber u_2 & = & \frac{1}{\sqrt{2}} \left(1,0,-1\right)^T,\\
\nonumber u_3 & = & \frac{1}{\sqrt{2}} \left(1,0,1\right)^T .
\end{eqnarray}
Finally, the eigenvectors of $C'$ are given by
\begin{eqnarray}
\nonumber w_1 & = & \frac{1}{\sqrt{2}} \left(1,0,1\right)^T ,\\
\nonumber w_2 & = & \frac{1}{\sqrt{2}} \left(1,0,-1\right)^T,\\
\nonumber w_3 & = & \left(0,1,0\right)^T .
\end{eqnarray}
The first two eigenvectors correspond to the eigenvalue $-1$ while the third one has the eigenvalue $1$. Notice that the second eigenvector can be chosen such that it is always the same. We parameterize the eigenvectors in such a way that they continuously change from $u_{1,3}$ to $w_{1,3}$ while remaining mutually orthogonal and normalized. This parametrization is given by
\begin{eqnarray}
\nonumber v_1(\rho) & = & \left(\frac{\rho}{\sqrt{2}},-\sqrt{1-\rho^2},\frac{\rho}{\sqrt{2}}\right)^T,\\
\nonumber v_2(\rho) & = & \frac{1}{\sqrt{2}}\left(1,0,-1\right)^T,\\
\nonumber v_3(\rho) & = & \left(\sqrt{\frac{1-\rho^2}{2}},\rho,\sqrt{\frac{1-\rho^2}{2}}\right)^T.
\end{eqnarray}
With these vectors we construct the following one-parameter set of coin operators
\begin{eqnarray}
\label{C:rho}
\nonumber C_2(\rho) & = & -P_1(\rho) - P_2(\rho) + P_3(\rho)\\
\nonumber & = &   \left(
              \begin{array}{ccc}
                -\rho^2 & \rho\sqrt{2(1-\rho^2)} & 1-\rho^2 \\
                \rho\sqrt{2(1-\rho^2)} & 2\rho^2-1 & \rho\sqrt{2(1-\rho^2)} \\
                1-\rho^2 & \rho\sqrt{2(1-\rho^2)} & -\rho^2 \\
              \end{array}
            \right).\\
\end{eqnarray}
The matrices $C$ and $C'$ correspond to the values $\rho = 0$ and $\rho = 1$, respectively. The Grover matrix is given by the coin parameter $\rho = \frac{1}{\sqrt{3}}$.

\begin{figure}[t!]
\begin{center}
\includegraphics[width=0.45\textwidth]{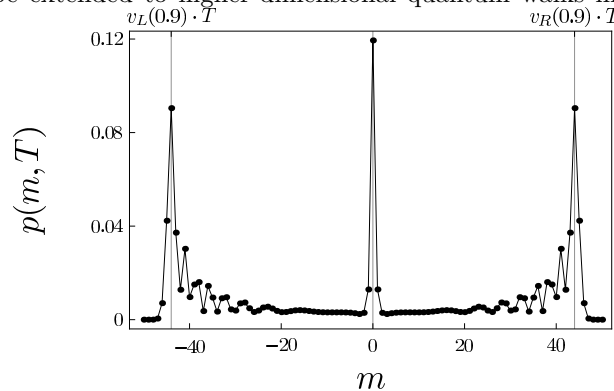}
\caption{The probability distribution of the three-state walk with the coin operator $C_2(\rho)$ after $T = 50$ steps. The initial coin state of the walk is $\psi_C = \frac{1}{\sqrt{2}}\left(1,0,1\right)$. As the coin parameter we have chosen $\rho=0.9$. This results in much faster spreading of the probability distribution. The peaks appear at the position $\pm\rho\cdot T = \pm 45$, in accordance with the analytical result (\ref{v:rho}).}
\label{fig4}
\end{center}
\end{figure}

We now show that the three-state quantum walks with the one-parameter family of coins (\ref{C:rho}) exhibits the localization effect and that the coin parameter $\rho$ directly determines the peak velocities. In order to prove this we analyze the dispersion relations
\begin{eqnarray}
\nonumber \omega_{1,2}(k,\rho) & = & \pm\arccos\left(\rho^2 - 1 - \rho^2\cos{k}\right),\\
\omega_3(k,\rho) & = & 0.
\end{eqnarray}
One of the frequencies is independent of the wavenumber, which guarantees the localization property of the corresponding one-parameter family of quantum walks with coin operator (\ref{C:rho}). Concerning the peak velocities, we have to determine for which wavenumber do the second derivatives of $\omega_{1,2}$ vanish. From their explicit form
$$
\frac{\partial^2\omega_{1,2}}{\partial k^2} = \pm\frac{\rho(\rho^2-1)\sqrt{1-\cos{k}}}{(2-\rho^2+\rho^2\cos{k})^{\frac{3}{2}}}
$$
we see that they are both equal to zero for $k_0=0$. Hence, the peak velocities are given by
\begin{eqnarray}
\label{v:rho}
\nonumber v_R(\rho) & = & \lim\limits_{k\rightarrow0^+}\frac{\partial\omega_2}{\partial k} = \rho,\\
v_L(\rho) & = & \lim\limits_{k\rightarrow0^+}\frac{\partial\omega_1}{\partial k} = -\rho .
\end{eqnarray}
Since $\rho$ can be varied from zero to one we can achieve faster spreading than for the Grover walk. We illustrate these results in Figure \ref{fig4}.

\section{Conclusions}
\label{sec5}

We have introduced two deformations of the Grover walk which preserve its localization nature. The coin parameters determine the velocities of the peaks in the probability distributions of the particle's position. The two families of walks differ in the achievable rate of spreading across the lattice. For the first one the upper limit is given by the original Grover walk. In the second case this limit on the peak velocity can be broken.

The presented construction of two sets of coins can be extended to higher-dimensional quantum walks in a straightforward way. In fact, the family of 2D quantum walks studied in \cite{konno:loc:2008} can be obtained by the deformation of the eigenvectors of the $4\times 4$ Grover matrix. Concerning the deformation based on the modification of the spectrum, one has to diagonalize the Grover matrix together with a permutation matrix which interchanges the displacements that mutually cancels each other. There is a unique eigenvector corresponding to eigenvalue $-1$ for the Grover matrix and eigenvalue 1 for the permutation. A construction similar to the one given in (\ref{C:phi}) yields a one-parameter set of coins preserving localization.

Our results show that localization effect can be found for a set of quantum walks. The presented construction is a step in a systematic classification of localizing quantum walks not only on a line but also in higher dimensions. It remains an open question if there exist coin operators outside the two families we have identified which also lead to localization.


\begin{acknowledgments}

We acknowledge the financial support from MSM 6840770039, MSMT LC06002 and SGS11/132/OHK4/2T/14.

\end{acknowledgments}


\bibliography{bibliography}{}

\end{document}